# Surface-dependent superconductivity and zero-energy end states in $UTe_2$

Fanbang Zheng[1,*], Wenhui Li[1,2,*,†], Dingsong Wu[3,*], Bin-Bin Zhang[4,*], Zhongzheng Yang[1], Shuai Qu[1,2], Ning Li[4], Chaofan Zhang[4,†], Gang Li[1,2,†], Yulin Chen[1,2,3,†], Shichao Yan[1,2,†]

[1]*State Key Laboratory of Quantum Functional Materials, School of Physical Science and Technology, ShanghaiTech University, Shanghai 201210, China*
[2]*ShanghaiTech Laboratory for Topological Physics, ShanghaiTech University, Shanghai 201210, China*
[3]*Department of Physics, University of Oxford, Oxford OX1 3PU, UK*
[4]*Nanhu Laser Laboratory, Changsha 410073, China*

*†Email: yanshch@shanghaitech.edu.cn; yulin.chen@physics.ox.ac.uk; ligang@shanghaitech.edu.cn; hjroland@163.com; liwh1@shanghaitech.edu.cn*

## Abstract

Surface-sensitive probes are central to elucidating the pairing mechanism in the candidate spin-triplet superconductor $UTe_2$, but their interpretation is complicated by multiple cleavage surfaces and surface reconstructions. Here, using ultralow-temperature scanning tunneling microscopy and spectroscopy, we identify two reconstructed $UTe_2$ surfaces with contrasting superconducting properties. The reconstructed (001) surface shows no clear superconducting gap feature, whereas the reconstructed (013) surface preserves a well-defined superconducting gap. On the (013) surface, magnetic-field-dependent differential-conductance imaging reveals single-flux-quantum vortices with highly anisotropic, Y-shaped vortex-core states. We further observe zero-bias peaks localized at step-edge chain termini, which can be suppressed by magnetic field without appreciable splitting. These observations provide signatures of zero-energy Andreev bound states associated with a sign-changing superconducting order parameter along the *a* axis of $UTe_2$. These findings establish surface termination and reconstruction as essential considerations in interpreting superconducting spectra and low-energy bound states in $UTe_2$.

## Introduction

The search for spin-triplet superconductivity remains a long-standing goal in condensed-matter physics [1,2]. $UTe_2$ has attracted intense interest as a candidate spin-triplet superconductor owing to its extremely large and anisotropic upper critical fields [3], magnetic-field-induced re-entrant superconducting phases [4], and anomalous Knight-shift behavior [5,6]. These properties make $UTe_2$ a promising platform for exploring spin-triplet pairing and topological superconductivity [7,8]. However, the nature of the superconducting order parameter in $UTe_2$ remains unresolved, with different experimental probes reporting conflicting results [9-15].

Surface-sensitive probes, including scanning tunneling microscopy and spectroscopy (STM/STS), surface-dependent Josephson tunneling and point-contact spectroscopy, have provided important information for identifying the pairing symmetry of $UTe_2$ [9,10,16,17]. However, the intrinsically three-dimensional crystal structure of $UTe_2$ means that cleavage can expose different crystallographic planes and terminations. Because surface termination and reconstruction can substantially alter the surface electronic structure, they may complicate the interpretation of surface-sensitive measurements. Direct atomic-scale characterization of distinct $UTe_2$ surfaces is therefore essential.

STM/STS provides simultaneous access to the surface atomic structure and local density of states. Previous STM/STS studies have focused mainly on the naturally cleaved $UTe_2$(0−11) surface, where superconducting gaps, chiral boundary modes, charge density wave and pair density wave orders, and vortex-core states have been reported [18-24]. In contrast, the atomic structures and electronic states of other cleavage surfaces remain largely unexplored.

Here, we combine STM/STS measurements and density functional theory calculations to identify two reconstructed $UTe_2$ surfaces corresponding to (001) and (013) orientations. The reconstructed (001) surface exhibits no clear superconducting gap feature, whereas the reconstructed (013) surface retains a well-defined superconducting gap. On the (013) surface, we observe highly anisotropic, Y-shaped vortex-core states and zero-bias peaks localized at the termini of step-edge chains. These results reveal a strong surface dependence of superconductivity and low-energy bound states in $UTe_2$, providing an atomic-scale reference for interpreting surface-sensitive measurements of this candidate spin-triplet superconductor.

**Results and discussion**

$UTe_2$ crystallizes in an orthorhombic structure containing U atoms and two inequivalent Te sites, Te(1) and Te(2) [Fig. 1(a)] [3]. For an ideal $UTe_2$(001) cleavage, Te(1)-, Te(2)- and U-terminated surfaces are possible. The U- and Te(1)-terminated surfaces are expected to form rectangular lattices with periodicities of 4.1 Å and 6.1 Å along the *a* and *b* axes [Fig. 1(b)], respectively, whereas an intact Te(2) termination has a shorter *b*-axis periodicity of 3.05 Å [upper panel in Fig. 1(c)]. The lattice parameters of $UTe_2$ used in the calculated model are listed in Supplemental Material [25], Table S1.

Before cleavage, we determined the (001) crystallographic orientation of the $UTe_2$ single crystal using X-ray Laue diffraction (see Supplemental Material [25], Fig. S1). Figure 1(e) shows a representative atomically resolved STM image of the cleaved $UTe_2$(001) surface. A zoomed-in image of the selected region in Fig. 1(e) is shown in Fig. 1(f), revealing the coexistence of chain-like and rectangular lattices within the same apparently flat area, with no discernible step edges. The height profile taken along the vertical green line in the rectangular-lattice region gives a nearest-neighbor spacing of approximately 4.1 Å [upper panel in Fig. 1(g)], consistent with the lattice parameter along the *a* axis [26]. From the horizontal line profiles in Figs. 1(f) and 1(g), the chain-like lattice shows a periodicity of approximately 3.1 Å. The minimum periodicity along the *b* axis of the Te(1)- and U-terminated lattices is 6.1 Å; neither can account for the 3.1 Å periodicity observed in the chain-like lattice. The observed periodicity is consistent with the expected periodicity of an intact Te(2)-terminated surface along the *b* axis. By contrast, the rectangular lattice shows a doubled periodicity of approximately 6.1 Å along the *b* axis [bottom panel in Fig. 1(g)]. This lattice is naturally assigned to a half-Te(2) surface, where every other surface Te(2) atom is absent [bottom panel in Fig. 1(c) and Supplemental Material [25], Fig. S2].

The coexistence of intact and half-occupied Te(2) lattices on the same flat surface suggests that cleavage occurs primarily through the breaking of U-Te(2) bonds. Depending on whether these bonds break on the same side or on alternating sides of the Te(2) layer, either a complete Te(2) lattice or a half-Te(2) lattice remains on the cleaved surface [Fig. 1(a)]. In some regions, the complete Te(2) lattice is also removed during cleavage, giving rise to a hole-like dark region, as indicated by the red arrow in Fig. 1(f). Moreover, we find that the remaining Te(2) atoms in the rectangular regions are shifted by half a lattice period along the *b* axis relative to the intact Te(2) lattice [bottom panel in Fig. 1(g)]. This reconstruction is reproduced by the simulated STM images for the intact and half-occupied Te(2) lattices [Fig. 1(d)], supporting the assignment of the rectangular regions to a reconstructed half-occupied Te(2) termination.

During cleavage, Te atoms in the Te(2) layer can either remain on the cleaved surface or be removed stochastically, leading to spatial variations in the appearance of the $UTe_2$(001) surface (Supplemental Material [25], Fig. S3). Correspondingly, differential-conductance ($dI/dV$) spectra acquired on different $UTe_2$(001) surfaces vary substantially, showing either weakly insulating behavior or a dip-like feature near the Fermi level [Fig. 1(h) and Supplemental Material [25], Fig. S4]. In all cases, however, no robust superconducting signatures are observed on the $UTe_2$ (001) surface. Although the blue spectrum in Fig. 1(h) exhibits a gap-like feature near the Fermi level, no clear superconducting vortices are observed in regions showing this spectral line shape under an applied magnetic field (Supplemental Material [25], Fig. S5). These results indicate that the $UTe_2$(001) surface has a reconstructed Te(2)-derived surface termination and exhibits strong disorder. The absence of a clear superconducting gap on the $UTe_2$(001) surface may be attributed to the disorder-induced pair breaking or reconstruction-related changes in the surface electronic structure.

In addition to the flat $UTe_2$(001) surface, we also observe a tilted surface on the same cleaved sample. A representative STM topographic image is shown in Fig. 2(a). The height profile yields a tilt angle of approximately 32° relative to the (001) plane, with typical values of $\theta \approx 34° \pm 3°$ across different regions (Supplemental Material [25], Fig. S6). To determine the intrinsic surface lattice, we subtract a linear background from the raw topographic image and correct the lateral dimensions along the tilt direction by a geometric factor of $1/\cos\theta$ (Supplemental Material [25], Fig. S7). The corrected atomically resolved image reveals a long-range ordered chain-like structure [Fig. 2(c)]. The nearest-neighbor spacing along the chains is approximately 4.1 Å, consistent with the crystallographic *a*-axis lattice constant of 4.1 Å for $UTe_2$, indicating that the atomic chains run along the *a* axis. Perpendicular to the chains, two neighboring chains form one periodic unit with a periodicity of about 11.5 Å [Fig. 2(c) and Supplemental Material [25], Fig. S8]. To distinguish the in-plane direction perpendicular to the *a* axis on the (013) surface from the *b* axis of bulk $UTe_2$, we denote it as $b^*$ [Fig. 2(c)].

The crystallographic orientation of this tilted surface can be assigned to the $UTe_2(013)$ plane based on the following observations. First, the calculated angle between the (013) and (001) planes of $UTe_2$ is approximately 37° [Fig. 2(b)], close to the experimentally measured tilt angle [Fig. 2(a)] . Second, as shown in [Fig. 2(b)], the ideal (013) cleavage mainly involves breaking U-Te(2) bonds and exposes alternating Te(1) and Te(2) crystallographic chains, naturally giving rise to a two-chain surface motif. The simulated STM image for the ideal (013) surface also shows a two-chain structure with a periodicity of 11.4 Å along the $b^*$ direction [upper panel in Fig. 2(e)]. The 4.1 Å periodicity along the chain indicates that the $a$ axis of $UTe_2$ lies within the tilted surface. Third, the measured step height across adjacent terraces is approximately 3.6 Å [Fig. 2(d)], close to the expected interplanar spacing of 3.7 Å for the $UTe_2(013)$ surface. Taken together, our experimentally measured tilt angle, in-plane lattice spacings, and interlayer spacing with standard crystallographic parameters allow us to assign this tilted surface to the $UTe_2(013)$ surface (Supplemental Material [25], Fig. S9).

However, we find that the observed tilted surface is reconstructed relative to the ideal (013) termination. For the ideal (013) surface, the two-chain motifs form a parallelogram-like surface lattice [upper panel in Fig. 2(e)]. In contrast, the experimental STM image shows that adjacent two-chain units are shifted by approximately half a lattice period along the $a$ axis, resulting in a rectangular lattice [Fig. 2(c)]. Although the simulated STM image of the reconstructed structure, in which the neighboring two-chain units on the (013) surface are displaced relative to one another by half a lattice constant along the $a$ axis, reproduces the observed surface pattern [bottom panel of Fig. 2(e)], our calculations indicate that this structure is metastable. This metastability suggests that cleavage-related effects or charge compensation associated with the polar (013) surface may help stabilize the reconstruction experimentally.

The reconstructed $UTe_2(013)$ surface hosts a clear superconducting gap. As shown in Fig. 2(f), site-resolved $\mathrm{d}I/\mathrm{d}V$ spectra taken at inequivalent atomic sites in Fig. 2(c) exhibit coherence peaks near ±0.32 mV, while the gap depth varies strongly across the surface (Supplemental Material [25], Fig. S10). The deepest gap appears between neighboring Te crystallographic chains [orange spectrum in Fig. 2(f)], whereas enhanced in-gap spectral weight is observed on Te(1) sites [green spectrum in Fig. 2(f)]. The deepest gap in the orange spectrum of Fig. 2(f) is significantly deeper than the typical gap depth measured on the $UTe_2(0{-}11)$ surface (Supplemental Material [25], Fig. S11). This site-dependent electronic structure likely reflects the distinct local bonding environments of Te(1) and Te(2) atoms created by cleavage-induced breaking of U-Te(2) bonds. Consistently, $\mathrm{d}I/\mathrm{d}V$ maps acquired at −0.32 mV and 0 mV show spatial modulations locked to the reconstructed lattice [Figs. 2(g) and 2(h)]. These results demonstrate that, in sharp contrast to the reconstructed (001) surface, the metastable reconstructed $UTe_2(013)$ surface preserves robust superconducting spectral signatures.

To further characterize superconductivity on the reconstructed $UTe_2(013)$ surface, we image magnetic-field-induced vortices using spatially resolved $\mathrm{d}I/\mathrm{d}V$ measurements. To enhance the vortex contrast, the zero-bias conductance map is normalized to the conductance at the coherence-peak energy, defined as $C(\boldsymbol{r}, 0\ \mathrm{mV}) = \frac{\mathrm{d}I/\mathrm{d}V(\boldsymbol{r}, 0\ \mathrm{mV})}{\mathrm{d}I/\mathrm{d}V(\boldsymbol{r}, 0.32\ \mathrm{mV})}$.

As shown in Fig. 3(a), the vortex core is highly anisotropic and elongated along the crystallographic $a$ axis. Linecut spectra along the $a$ and $b^*$ directions reveal the spatial evolution of the vortex-core states [Figs. 3(b) and 3(c)]. At the vortex center, the spectrum exhibits a pronounced zero-bias conductance peak [Fig. 3(d)]. Moving away from the vortex-core center along the $a$ axis, this zero-bias peak persists over a distance of approximately 12 nm before splitting into two finite-energy peaks, forming a characteristic Y-shaped spatial evolution. Similar Y-shaped vortex-core spectra have been observed in several unconventional or topological superconducting systems [27,28]. Along the $b^*$ direction, the zero-energy vortex-core state decays over a shorter length scale of about 5 nm before showing a similar splitting behavior (Supplemental Material [25], Fig. S12). These results demonstrate a strong in-plane anisotropy of the vortex-core bound states on the $UTe_2$(013) surface. In addition, spectra taken on opposite sides of the vortex core show noticeable left-right asymmetry [Fig. 3(e)], reminiscent of the symmetry-broken vortex-core spectra reported on the $UTe_2$(0−11) surface. However, the zero-energy vortex-core state observed on the $UTe_2$(0−11) surface does not show clear energy splitting along the $a$ axis (Supplemental Material [25], Fig. S13). The difference may reflect the distinct electronic states and superconducting gap structures on the two surface terminations.

We next examine the magnetic-field dependence of the vortices. As the out-of-plane magnetic field increases, the vortex density increases accordingly [Figs. 3(g)–3(i)]. The extracted vortex number follows $N = B \cdot S/\Phi_0$, where $S$ is the imaging area and $\Phi_0 = 2.07\times10^{-15}$ Wb is the superconducting flux quantum [Figs. 3(j) and Supplemental Material [25], Fig. S14], confirming that the observed vortices are single-flux-quantum vortices. Finally, the vortex-core spectrum evolves systematically with magnetic field: the zero-bias peak gradually weakens from 1 T to 5 T without clear splitting [Fig. 3(f)]. This suppression may result from enhanced intervortex coupling as the intervortex distance decreases at higher fields [29]. When the associated energy scale is smaller than the intrinsic and experimental linewidth, this hybridization may not produce resolvable peak splitting [29].

Motivated by the previously reported edge states on the $UTe_2$(0−11) surface [18], we search for boundary states associated with step edges on the superconducting $UTe_2$(013) surface. Figure 4(a) shows a representative STM topographic image containing multiple step-edge chains running along the $a$ axis, each consisting of two rows of Te atoms. The spectrum acquired on a step edge exhibits an asymmetric in-gap resonance, with a peak at negative bias and a corresponding dip at positive bias [purple spectrum in Fig. 4(c)], resembling the edge-state feature reported for one type of step edge on the $UTe_2$(0−11) surface [18]. More strikingly, pronounced zero-bias peaks are observed at the termini of the step-edge chains [Fig. 4(c)]. Their spatial localization is directly visualized in the zero-bias $\mathrm{d}I/\mathrm{d}V$ map, where the enhanced zero-bias conductance is concentrated at the step-edge chain ends [Fig. 4(b)]. Upon applying a magnetic field, the zero-bias peak is gradually suppressed without appreciable splitting up to 5 T [Fig. 4(e)]. Since the superconducting gap is simultaneously weakened by the magnetic field, this behavior indicates that the zero-energy end states are closely related to superconductivity-induced in-gap quasiparticle states rather than purely structural or normal-state defect resonances.

The localization of the zero-bias peaks at the step-edge chain termini and their magnetic-field suppression are consistent with zero-energy Andreev bound states associated with a sign-changing superconducting order parameter. As schematically illustrated in Fig. 4(d), for an odd-parity pairing state with a dominant $p_a$-like component, the superconducting order parameter changes sign upon reversal of the $a$-axis momentum component from $k_a$ to $-k_a$. Quasiparticle reflection at a step-edge chain termination reverses $k_a$, thereby changing the sign of the order parameter. The resulting intrinsic phase shift can generate zero-energy Andreev bound states [30,31], providing a possible explanation for the observed zero-energy end states. In this scenario, the superconducting gap of $UTe_2$ would depend predominantly on $k_a$, with gap nodes located away from the $a$ axis. Such a gap structure in $UTe_2$ would therefore disfavor a pure $B_{3u}$ pairing state, which has point nodes along the $a$ axis [9,32].

## Conclusion

In summary, STM/STS measurements reveal two reconstructed cleavage surfaces of $UTe_2$, namely the (001) and (013) surfaces, with markedly different electronic properties. The reconstructed (001) surface exhibits no robust superconducting signatures, whereas the reconstructed (013) surface preserves a well-defined superconducting gap. On the (013) surface, we observe single-flux-quantum vortices with highly anisotropic, Y-shaped vortex-core states, as well as zero-bias peaks localized at the termini of step-edge chains along the $a$ axis. These end-localized states provide signatures of superconductivity-related boundary modes associated with chain terminations. Our findings highlight the crucial role of surface termination and reconstruction in $UTe_2$ and provide an atomic-scale reference for interpreting surface-sensitive probes of its unconventional superconductivity.

## Acknowledgements

S.Y. acknowledges the financial support from the National Natural Science Foundation of China (Grant No. 12674077). S.Y. and G.L. acknowledge the financial support from the National Key Research and Development Program of China (Grant No. 2022YFA1402703) and the start-up funding from ShanghaiTech University. C.Z. acknowledges the financial support from Outstanding Young Researcher Scheme of Hunan Province (Grant No. 2023JJ10051). W.L. acknowledges the financial support from the National Natural Science Foundation of China (Grant No. 12404222).

## Figures and figure captions

### Figure 1

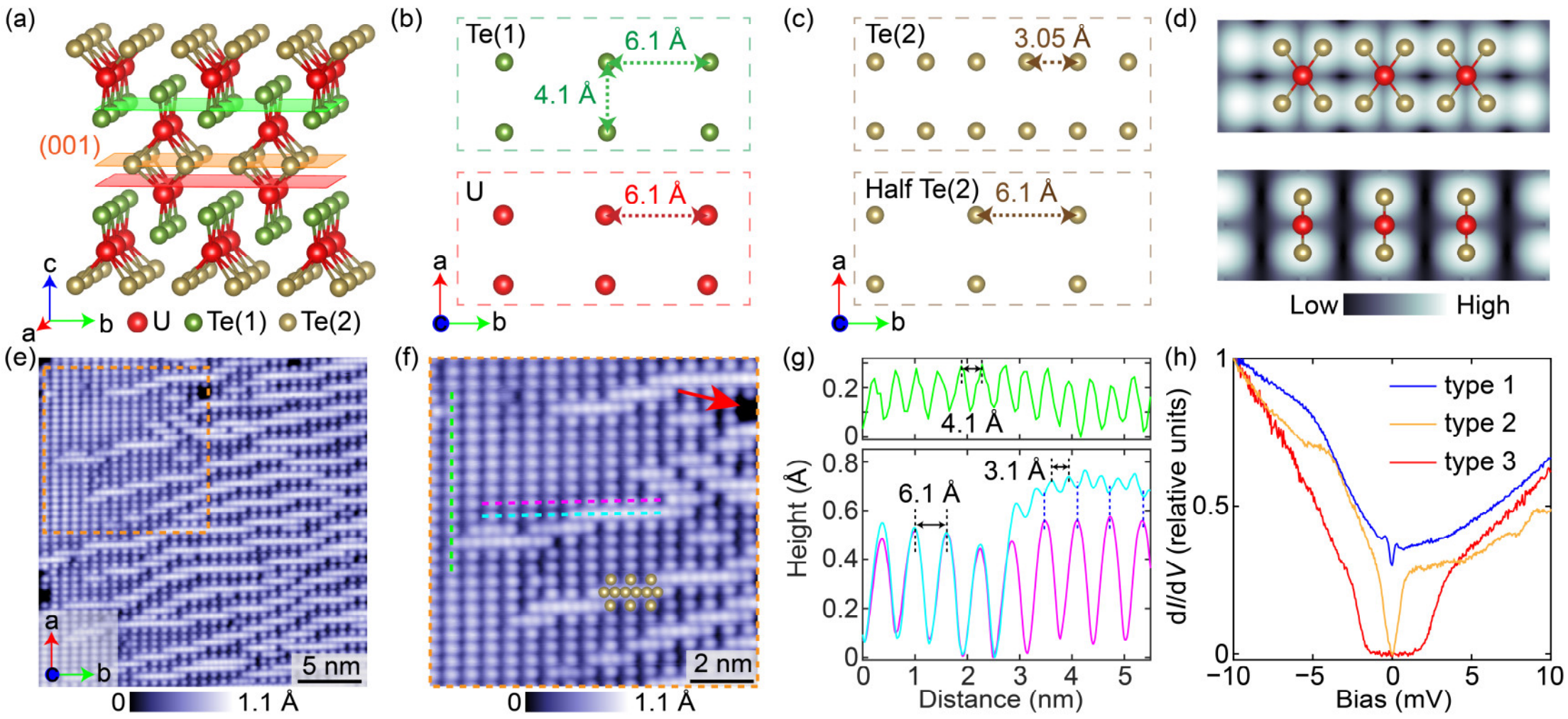

FIG. 1. Atomic and electronic structure of the $UTe_2$(001) surface. (a) Crystal structure of $UTe_2$. Green, orange and red shaded planes mark possible (001) cleavage planes exposing Te(1), Te(2) and U layers, respectively. (b) Top-view schematics of Te(1)- and U-terminated (001) surfaces, both with 4.1 Å and 6.1 Å periodicities along the *a* and *b* axes. (c) Schematics of intact and half-Te(2)-terminated (001) surfaces. The intact Te(2) surface has a 3.05 Å periodicity along the *b* axis, whereas the half-Te(2) surface shows a doubled 6.1 Å periodicity. (d) Simulated STM images of the intact and half-occupied Te(2) terminations, respectively. The bottom schematic shows that the remaining Te(2) atoms in the half-occupied termination are shifted by half a lattice period along the *b* axis relative to the intact Te(2) lattice. (e) STM topographic image of the $UTe_2$(001) surface showing coexisting chain-like and rectangular lattices ($V_{set}$ = −150 mV, $I_{set}$ = 150 pA). (f) Zoomed-in image of the region marked by the orange dashed box in (e) ($V_{set}$ = −150 mV, $I_{set}$ = 100 pA). The red arrow indicates a hole-like dark region. (g) Height profiles along the colored dashed lines in (f), showing periodicities of 4.1 Å along the *a* axis and 3.1 Å or 6.1 Å along the *b* axis. (h) Three representative d$I$/d$V$ spectra acquired at different locations on the (001) surface indicated in Fig. S4. These spectra are normalized to the differential conductance at −10 mV.

**Figure 2**

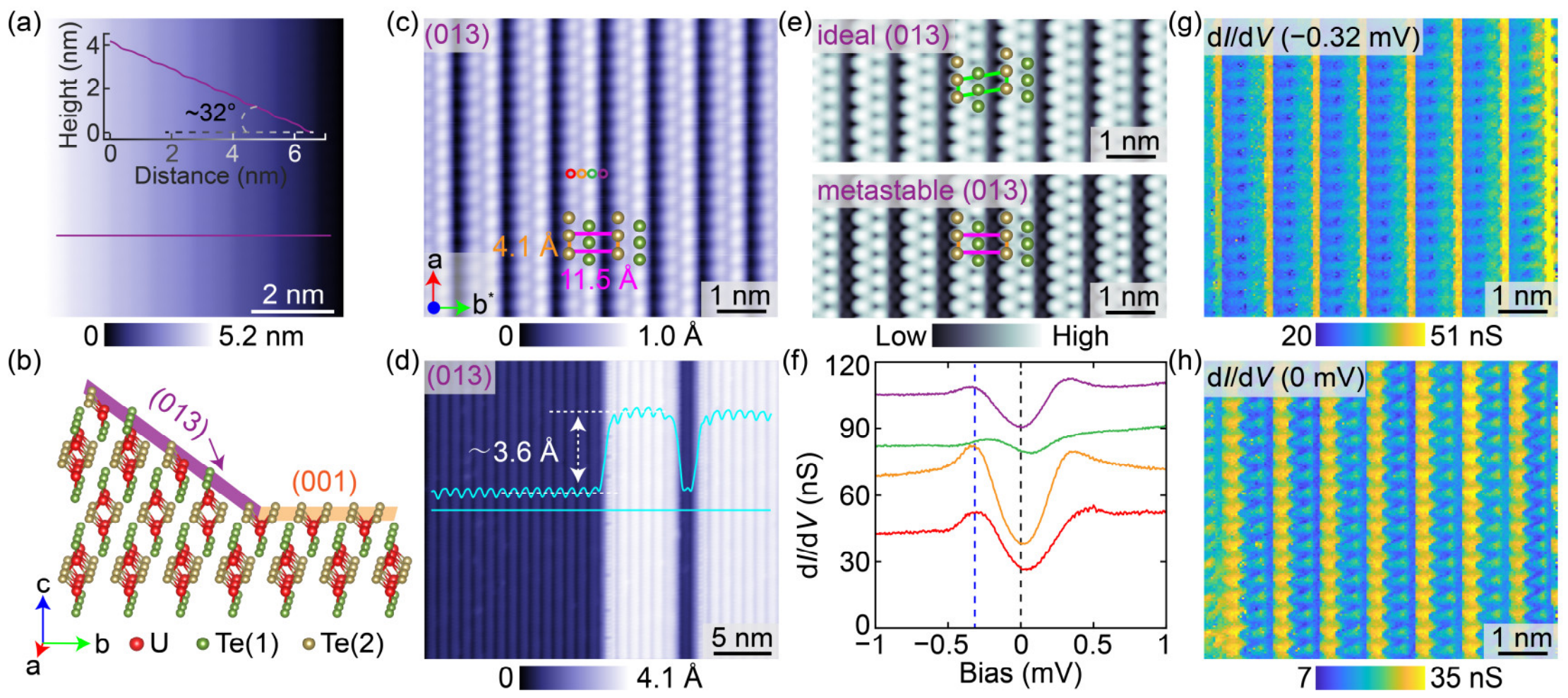

FIG. 2. Atomic and electronic structure of the $UTe_2$(013) surface. (a) STM topographic image of a tilted $UTe_2$ terrace. The inset shows a height profile along the purple line, giving an inclination angle of approximately 32° relative to the (001) plane. (b) Crystal-structure schematic showing the proposed (001) and (013) cleavage planes, marked in orange and purple, respectively. (c) Atomically resolved STM image of the $UTe_2$(013) surface overlaid with the proposed surface model. Orange and purple markers indicate intra-chain and inter-chain spacings between neighboring Te chains, respectively ($V_{set}$ = −50 mV, $I_{set}$ = 50 pA). (d) STM image across adjacent (013) terraces, with a height profile along the cyan line showing a step height of ~3.6 Å ($V_{set}$ = −90 mV, $I_{set}$ = 50 pA). (e) Simulated STM images of the ideal and reconstructed metastable (013) surfaces. The reconstructed metastable surface reproduces the experimentally observed double-chain surface structure. (f) Representative $dI/dV$ spectra acquired at inequivalent atomic sites marked in (c) ($V_{set}$ = −10 mV, $I_{set}$ = 500 pA). The spectra are vertically offset by 22 nS for clarity. (g),(h) $dI/dV$ maps acquired over the same field of view as (c) at −0.32 mV and 0 mV, respectively.

## Figure 3

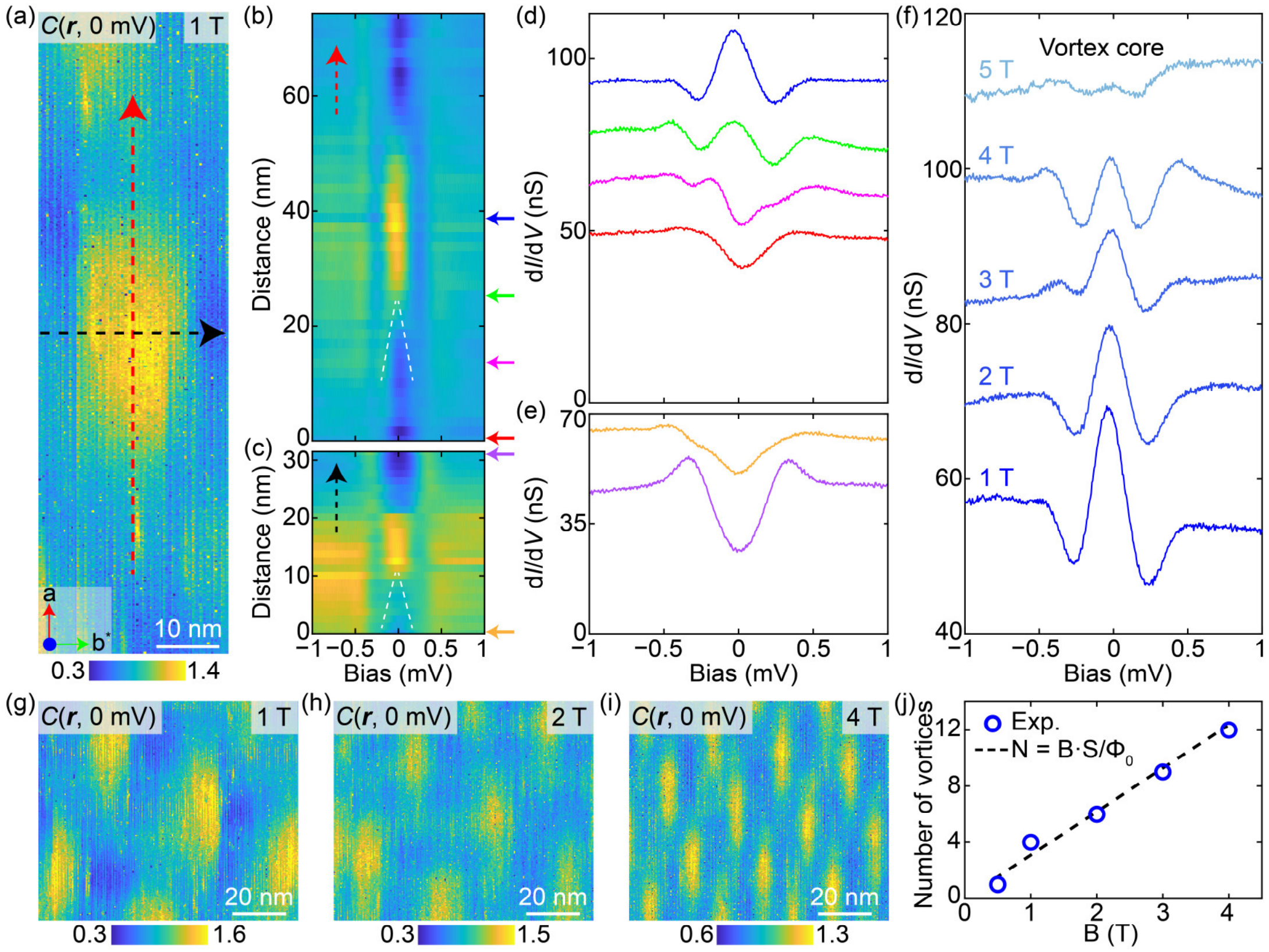

FIG. 3. Vortex-core states on the $UTe_2$(013) surface. (a) Normalized zero-bias conductance map, $C(\boldsymbol{r}, 0\ \text{mV})$, of vortices on the $UTe_2$(013) surface under a perpendicular magnetic field of 1 T. (b),(c) Spatially resolved d$I$/d$V$ spectra taken along the red and black arrows in (a), respectively. Dashed lines trace the splitting of the zero-bias vortex-core state into finite-energy peaks away from the vortex center. (d),(e) Representative d$I$/d$V$ spectra extracted from selected positions in (b) and (c), marked by colored arrows. (f) Magnetic-field evolution of the vortex-core spectrum. (g)-(i) Normalized zero-bias conductance maps acquired at different perpendicular magnetic fields, showing an increasing vortex density. (j) Vortex number as a function of magnetic field. The dashed line shows the expected relation $N = B \cdot S/\Phi_0$, where $B$ is the applied field, $S$ is the imaging area and $\Phi_0 = 2.07 \times 10^{-15}$ Wb is the superconducting flux quantum. All spectra and maps were acquired with $V_{set} = 10$ mV and $I_{set} = 500$ pA. The spectra in (d) are vertically offset by 10 nS, 10 nS and 23 nS for clarity. The spectra in (e) are vertically offset by 20 nS for clarity. The spectra in (f) are vertically offset by 30 nS, 10 nS, 3 nS and 22 nS for clarity.

## Figure 4

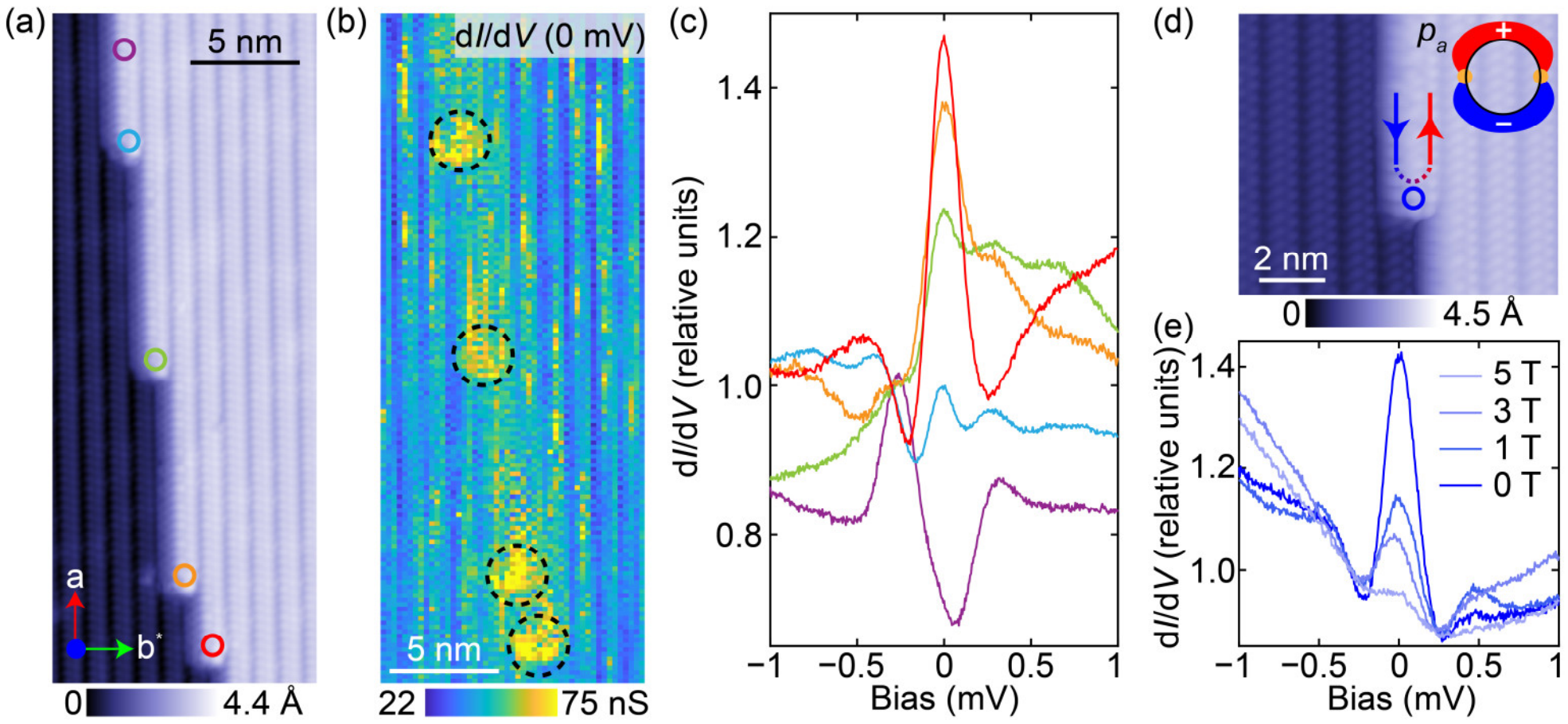

FIG. 4. Zero-energy end states at step-edge chain termini on the $UTe_2(013)$ surface. (a) STM topographic image of the $UTe_2(013)$ surface showing step-edge chains running along the *a* axis ($V_{set} = -90$ mV, $I_{set} = 50$ pA). Colored circles indicate the positions of the spectra in (c). (b) Zero-bias $dI/dV$ map acquired over the same field of view as (a), showing enhanced conductance at chain termini ($V_{set} = -5$ mV, $I_{set} = 500$ pA). Dashed circles highlight the enhanced zero-bias conductance at the chain ends. (c) $dI/dV$ spectra acquired at the positions marked in (a)**.** The orange spectrum is acquired with $V_{set} = -5$ mV and $I_{set} = 500$ pA, whereas the other spectra are acquired at $V_{set} = -3$ mV and $I_{set} = 700$ pA. (d) STM topographic image showing the chain terminus used for the field-dependent measurements in (e) ($V_{set} = 90$ mV, $I_{set} = 50$ pA). The blue circle marks the spectroscopic measurement position. The arrows schematically illustrate Andreev reflection at the chain termination. The inset shows the sign-changing gap structure of a $p_a$-wave pairing state, with opposite signs of the pair potential for $k_a$ and $-k_a$. (e) Magnetic-field dependence of the zero-bias peak at the chain terminus ($V_{set} = 10$ mV, $I_{set} = 500$ pA). The spectra in (c) and (e) are normalized to the differential conductance at −0.3 mV.